\documentclass[aps,nofootinbib,superscriptaddress,10pt,floatfix]{revtex4}
\normalsize
\usepackage{mathrsfs}
\usepackage[scr=boondox]{mathalpha}  
\usepackage{graphicx}
\usepackage{amsmath,amssymb}
\usepackage{braket}
\usepackage{subfigure}
\usepackage{subfloat}
\usepackage{float}
\usepackage{mciteplus}
\usepackage[normalem]{ulem}
\usepackage[usenames]{color}
\usepackage[normalem]{ulem}
\usepackage{comment}
\usepackage{ifthen} 
\usepackage{fp,calc}
\usepackage{lastpage}
\usepackage{setspace}
\usepackage[pdftex]{epsfig}
\DeclareGraphicsExtensions{.jpg,.mps,.pdf,.png}
\usepackage{cellspace}
\usepackage{booktabs}
\usepackage{latexsym}
\usepackage{graphics}
\usepackage{fancyhdr}
\usepackage{adjustbox}
\usepackage[english]{babel}
\usepackage[svgnames]{xcolor}
\usepackage{enumitem}
\usepackage{tabularx}

\newcommand{\bn}{\mathbf{n}}

\newcommand{\dd}{{\rm d}}
\newcommand{\E}{{\rm e}}

\def\beq{\begin{equation}}
\def\eeq{\end{equation}}
\newcommand{\rbr}[1]{\left(#1\right)}
\newcommand{\sbr}[1]{\left[#1\right]}
\newcommand{\angbr}[1]{\left\langle#1\right\rangle}
\newcommand{\pg}{\mathfrak{p}_g}
\newcommand{\cond}{_{h\vert\rho_g}}
\newcommand{\nnn}{\nonumber \\}
\newcommand{\st}{{\mathrm{st.}}}
\newcommand{\clust}{{\mathrm{clust.}}}

\newcommand{\be}{\begin{equation}}
\newcommand{\ee}{\end{equation}}

\usepackage{float}
\usepackage{graphicx}

\newboolean{articletitles}
\setboolean{articletitles}{true} 
\newboolean{allauthors}
\setboolean{allauthors}{false} 

\mciteErrorOnUnknownfalse
\usepackage[colorlinks=true, allcolors=blue]{hyperref}

\definecolor{Mgreen}{rgb}{0.1, 0.69,0.16}

\begin{document}
 
\title{The impact of large-scale galaxy clustering on the variance of the Hellings-Downs correlation: theoretical framework}

\author{Nastassia Grimm}
\email[]{nastassia.grimm@unige.ch}
\affiliation{D\'epartement de Physique Th\'eorique and Center for Astroparticle Physics, Universit\'e de Gen\`eve, Quai E. Ansermet 24, CH-1211 Geneve 4, Switzerland}

\author{Martin Pijnenburg}
\email[]{martin.pijnenburg@unige.ch}
\affiliation{D\'epartement de Physique Th\'eorique and Center for Astroparticle Physics, Universit\'e de Gen\`eve, Quai E. Ansermet 24, CH-1211 Geneve 4, Switzerland}

\author{Giulia Cusin}
\email[]{giulia.cusin@iap.fr}
\affiliation{Institut d'Astrophysique de Paris, UMR-7095 du CNRS, Paris, France}
\affiliation{D\'epartement de Physique Th\'eorique and Center for Astroparticle Physics, Universit\'e de Gen\`eve, Quai E. Ansermet 24, CH-1211 Geneve 4, Switzerland}

\author{Camille Bonvin}
\email[]{camille.bonvin@unige.ch}
\affiliation{D\'epartement de Physique Th\'eorique and Center for Astroparticle Physics, Universit\'e de Gen\`eve, Quai E. Ansermet 24, CH-1211 Geneve 4, Switzerland}

\date{\today}

\begin{abstract}

While pulsar timing array experiments have recently found evidence for the existence of a stochastic gravitational wave (GW) background, its origin is still unclear. If this background is of astrophysical origin, we expect the distribution of GW sources to follow the one of galaxies. Since galaxies are not perfectly isotropically distributed at large scales, but follow the cosmological large-scale structure, this would lead to an intrinsic anisotropy in the distribution of GW sources. In this work, we develop a formalism to account for this anisotropy, by considering a Gaussian ensemble of sources in each realization of the universe and then taking ensemble averages over all such realizations. We find that large-scale galaxy clustering has no impact on the expectation value of pulsar timing residual correlations, described by the Hellings-Downs curve. However, it introduces a new contribution to the variance of the Hellings-Downs correlation. Hence, the anisotropic distribution of sources contributes to the amount by which the measurements of pulsar timing residual correlations, in our single realization of the universe, may differ from the Hellings-Downs curve. 

\end{abstract}

\maketitle


\section{Introduction}

Recently, pulsar timing arrays (PTAs) were used to deliver evidence for the existence of a stochastic gravitational wave (GW) background~\cite{NANOGrav:2023gor, Reardon:2023gzh, EPTA:2023sfo, Xu:2023wog}. This detection utilizes the angular correlation of timing residuals, i.e.~small perturbations in the arrival times of radio pulses caused by the passing of GWs.  
The origin of the GW background is still unclear, and it might be of both astrophysical or cosmological nature (see e.g.~Refs.~\cite{Shannon:2015ect, Chen:2019xse, NANOGrav:2021flc, NANOGrav:2023hvm, Vagnozzi:2023lwo}). One possibility to distinguish between these scenarios is to evaluate whether galaxy clustering has an impact on the angular correlation of timing residuals: if sources contributing to the GW background are of astrophysical origin, we expect their distribution to follow the distribution of galaxies. Galaxies are however not perfectly isotropically distributed on the sky: they are grouped in clusters, that are themselves arranged into a large-scale structure. Anisotropies in the galaxy distribution have been observed by various surveys, leading to correlations between galaxy positions up to very large cosmological scales, see e.g.\ Ref.~\cite{Blake_2011,Howlett:2014opa,Pezzotta:2016gbo,Alam2016:1607.03155v1,eBOSS:2020yzd}. Therefore, any GW background of astrophysical origin is expected to be as well anisotropic: an intrinsic source of anisotropy is due to clustering (the distribution of GW sources follows the one of the large-scale structure), a secondary one is due to the fact that once a GW signal is emitted, it feels the effect of the gravitational potential of structures \cite{Cusin:2017fwz, Cusin:2017mjm, Cusin:2018rsq}. {These} secondary anisotropies are expected to be subdominant on all scales \cite{Pitrou:2019rjz}, hence we will focus here on the effect of large-scale galaxy clustering.  

Usually, the correlation of pulsar timing residuals due to the passing of GWs, as a function of the pulsar pair separation on the sky, is assumed to be described by the Hellings and Downs (HD) curve~\cite{Hellings:1983fr}. Recent literature has investigated to which extent this curve is precise in the description of correlations of timing residuals~\cite{Allen:2022dzg, Allen:2022ksj, Romano:2023zhb, Allen:2024rqk, Bernardo:2022xzl}. It has been found that, in practice, PTA observations are not expected to exactly recover the HD curve in any realistic models of the GW background. For instance, the interference of GW sources would cause a departure from this idealized curve~\cite{Allen:2022bjz, Allen:2022dzg}. In all these studies, the fact that sources are clustered is systematically neglected.\footnote{The possibility of anisotropies in the distribution of GW sources in the PTA band has been pointed out in previous work, however either not in the context of the HD correlation \cite{Allen:1996gp,Cusin:2017fwz, Cusin:2017mjm, Cusin:2018rsq, Cusin:2018avf, Cusin:2019jpv, Cusin:2019jhg, Pitrou:2019rjz, Jenkins:2019nks, Alonso:2020mva, renzini2022}, or without targeting the impact on its variance \cite{Mingarelli:2013dsa, Taylor:2013esa, Gair:2014rwa, Ali-Haimoud:2016mbv}.}

In this work, we present, for the first time, a framework to account for the impact of the cosmological large-scale structure on the HD correlation.\footnote{Note that in this manuscript, we will use the notion \textit{HD correlation} to denote the observed correlation of pulsar timing delays, while \textit{HD curve} denotes the theoretical prediction made by Hellings \& Downs in 1983~\cite{Hellings:1983fr}.} This method, novel in the context of PTAs, is based on a two-step averaging approach: first, we describe a Gaussian ensemble of GW sources in a realization of the universe with a given distribution of galaxies. Second, to determine the impact of the galaxy distribution on the HD correlation, we study how it varies not only from ensemble to ensemble of GW sources, but also from realization to realization of the galaxy distribution. Mathematically, the latter is done by taking a second ensemble average over all such realizations.  
We find that the mean of the HD correlation itself is not affected, which means that anisotropies in the distribution of sources do not generate a systematic bias in the HD correlation. In other words, if we were able to measure the correlations of pulsar timing residuals in many realizations of the universe, we would in average recover the HD curve predicted in Ref.~\cite{Hellings:1983fr} for the case of isotropically distributed sources. However, in practice we have only access to one such realization. 
To account for this, we calculate the variance of the HD correlation due to the large-scale structure of the Universe, which tells us by how much the measurements in our specific realization of the universe can differ from the mean. In this paper we find that, indeed, large-scale galaxy clustering leads to a new contribution to the HD variance, which needs to be carefully evaluated in light of future observations. {Numerical results for specific population models of supermassive black holes are presented in a companion paper~\cite{Grimm:2024hgi}.} 

In addition to the variance due the clustering of sources, there is as well a variance due to shot noise, i.e.\ due to the fact that GW sources are not a continuous field, but  rather have a discrete Poisson distribution (see e.g.\ Ref.~\cite{Jenkins:2019nks}). In this paper we do not take the shot noise contribution into account (see Ref.~\cite{Allen:2024mtn} for a derivation of this contribution for PTAs), and we concentrate only on the modeling of the clustering variance. Consequently, we refer to an \emph{isotropic distribution} of sources when no clustering is present, and an \emph{anisotropic distribution} of sources when clustering is accounted for. This use of terminology, although being common in literature, is not fully accurate due to shot noise: since sources have a discrete nature, they will never have a perfectly isotropic distribution, even in the absence of clustering.

The remainder of this work is structured as follows: In section~\ref{sec:Ensemble_averages}, we first revise basic notations in the case of an isotropic distribution of GW sources, and then introduce galaxy clustering and discuss the respective ensemble averaging procedure. We present our calculation of the variance of the HD correlation due to large-scale clustering in section~\ref{Sec:HDvariance}, with our main result given in Eq.~\eqref{Eq:2ndmom_rho_ab_clustering}. We conclude in section~\ref{sec:conclusion}, and comment on the positivity of this novel contribution to the variance in Appendix~\ref{sec:Positivity}.

\section{Ensemble averages of GW signals} \label{sec:Ensemble_averages}

Before calculating the impact of large-scale structure on the HD correlation, we first determine how it impacts the two-point function of the GW amplitude. In the following, we first discuss the different layers of stochasticity present in our analyses, then discuss the GW signal for an isotropic distribution of sources, and subsequently introduce the impact of galaxy clustering.

\subsection{Two different layers of stochasticity} \label{sec:two_stochasticities}
The GW signal is a stochastic quantity, first due to the fact that it arises from an (unknown) superposition of waves, and second because we do not know the spatial distribution of sources.  As a consequence, the HD correlation is also a statistical quantity. The aim of this paper is to compute the mean and variance of the HD correlation accounting for both sources of stochasticity. We introduce therefore two types of averages, laying the foundation for the subsequent sections:
\begin{enumerate}[label=(\roman*)]
\item \textit{Ensemble average over GW realizations.} The GW signal is treated as a stochastic gravitational wave background (SGWB). It arises indeed from the superposition of a large number of sources, emitting incoherently at distinct frequencies. The total strain in a given direction is stochastic since we do not know the details of the distribution of sources emitting GWs during the observing time $T$ (e.g.\ we do not know at which stage of the inspiral a given source is, and this reflects in a stochasticity of both amplitude and phase; orbital planes are also randomly oriented, hence the GW polarization is random; see also Ref.~\cite{Maggiore:1900zz} for a description of the stochastic background). We can therefore not predict the value of the strain, but we can predict its 2-point function (since we know the statistical properties of the sources), determined by the spectral density.
\item \textit{Ensemble average
over stochastic initial conditions of cosmological variables, in particular the galaxy density field.} The stochasticity of
these variables is inherited from their quantum
origin during inflation. Since the distribution of GW sources follows the distribution of matter, the total strain obtained by summing the contribution from different sources along the line of sight inherits the stochasticity of the matter distribution. We note that this average (ii) is the usual
ensemble average used in cosmology to compute
correlation functions and angular power spectra of cosmological observables. We also emphasize that if a GW observable is a stochastic quantity, then the average (i) over GW realizations alone still maintains the stochasticity of the matter distribution. 
\end{enumerate}
We emphasize again that, in practice, we only have observational access to one realization of the Universe, i.e.~only one realization of the SGWB and of the galaxy density field. Therefore, for any observable, it is important to know its variance in addition to its theoretical expectation value: the variance provides a measure by how much we expect the observation and the theoretical expectation value to differ.

\subsection{Isotropic distribution of sources} \label{sec:noclust}

We start by summarizing the basic concepts and notations 
(see e.g.~Ref.~\cite{Maggiore:1900zz} for a standard reference) necessary to describe the GW signal in absence of anisotropies in the galaxy density (i.e., the average (ii) described above plays no role yet). We work in units where the speed of light is given by $c=1$ for convenience. We consider a homogeneous and isotropic universe without galaxy clustering and a stochastic GW background, causing a perturbation to the space-time metric:
\begin{equation}
\mathrm ds^2=-\mathrm dt^2+\mathrm d\mathbf{x}^2+h_{ij}(t,\mathbf{x})\mathrm dx^i\mathrm dx^j\,.
\end{equation}
We work in the transverse-traceless synchronous gauge, where the quantity $h_{ij}$ can be written, using the plane-wave expansion, as
\begin{equation}
h_{ij}(t,\mathbf{x})=\sum_{A=+, \times}  \int\mathrm df\int \mathrm d\bn\, h_A(f, \bn)\E^{2\pi if(t-\boldsymbol{n}\cdot \mathbf{x})}e^A_{ij}(\bn)\,,
\end{equation}
where $e_{ij}^A(\bn)$ is the polarization tensor and $h_A(f,\bn)$ denotes the amplitude of the polarization state $A$ at frequency $f$ and propagating in direction $\bn$ (hence originating from direction $-\bn$). The amplitudes are complex functions satisfying $h_A(-f,\bn)=h_A^\ast(f,\bn)$. Moreover, by definition of a stochastic background the amplitudes have zero mean: $\angbr{h_A(f,\bn)}=0$. The expectation value of the product of two $h_A(f,\bn)$ is however non-zero and given by
\begin{equation}
\left\langle h_A^\ast(f,\mathbf{n})h_B(f',\mathbf{n'})\right\rangle =\frac 12 \delta_{AB}\,\delta(f-f')\bar S_h(f)\frac{\delta^2(\mathbf{n},\mathbf{n'})}{4\pi}\,. \label{eq:hh_noclust}
\end{equation} 
The delta functions account for the fact that only GWs with the same polarization state, at the same frequency and coming from the same direction are correlated \cite{Cusin:2017mjm}. In other words, at each unique polarization state and unique value of $f$ and $\bn$, the quantity $h_A(f,\bn)$ is a Gaussian random variable with variance determined by the spectral density $\bar S_h(f)$.\footnote{This spectral density is related to the isotropic GW energy parameter $\bar{\Omega}_\text{GW}$ via $\bar{\Omega}_\text{GW}= \pi f^3\bar{S}_h/(2G\rho_c) $, with $\rho_c$ the critical energy density of the Universe \cite{Maggiore:1900zz}.} 
For a stochastic background due to astrophysical sources, the spectral density  $\bar S_h(f)$ appearing in Eq.\,(\ref{eq:hh_noclust}) is related to the number of sources as well as to their characteristics. It depends on frequency, but by construction it does not depend on directions $\bn$ given that we assume an isotropic distribution of GW sources. Thus, while $h_A(f,\bn)$ generally depends on direction $\bn$ due to the stochasticity of the GW background, its variance does not unless other forms of anisotropy are present in the Universe.

The expectation value given in Eq.~\eqref{eq:hh_noclust} occurs in standard calculations leading to the HD curve, i.e.~in the calculation of the mean of timing residual correlations, for a Gaussian ensemble of GW sources. As a next step, we want to extend Eq.~\eqref{eq:hh_noclust} to account for the fact that GW sources follow the large-scale structure of the Universe, hence have an anisotropic distribution. This will require taking the ensemble averaging on the left-hand side of Eq.~\eqref{eq:hh_noclust} in two steps, in order to distinguish the two different layers of stochasticity described in section~\ref{sec:two_stochasticities}.

\subsection{Galaxy clustering}
 
In an inhomogeneous universe, the galaxy density distribution can be treated as a random field (see average (ii) in section~\ref{sec:two_stochasticities}), directly linked to the primordial fluctuations generated by inflation. We denote by $\mathfrak{p}_g(\bn,z)$ this random field at redshift $z$ and spatial coordinates $\mathbf{x}=- r_z\bn$, with $r_z$ being the comoving distance\footnote{The comoving distance is related to redshift as $r_z=\int_0^z{\mathrm dz'}\,{H(z')^{-1}}$, where $H(z)$ is the Hubble parameter. The negative sign in the spatial coordinates $\mathbf{x}=-r_z\bn$ is introduced for notational convenience in subsequent sections, where galaxy density will be related to GWs propagating in direction $\bn$, i.e.~originating from direction $-\bn$.}. Our Universe is a particular realization of this random field, that we denote by $\rho_g(\bn,z)$. Assuming that inflation generates statistically isotropic fluctuations (i.e.~no statistically preferred direction is present in the Universe, which means that statistical expectation values are preserved under rotation on the sky, see e.g.\ Refs.~\cite{Armendariz-Picon:2005lfa, Planck:2015igc, Hajian:2003qq}), the mean of $\mathfrak{p}_g(\bn,z)$ over all possible realizations does not depend on direction: 
\beq
\angbr{\pg(\bn,z)}=\bar\rho_g(z)\,.
\eeq 
Due to the ergodic theorem, this mean also represents the average distribution of galaxies over all directions $\bn$ in one realization $\rho_g(\bn,z)$.\footnote{In the context of cosmological large-scale structure, the ergodic theorem usually refers to replacing ensemble averages over different realizations of the universe by spatial averages over different regions of the sky, see e.g.\ Ref.~\cite{2012reco.book.....E}.} 
The covariance of $\pg(\bn,z)$ is given by
\beq
\angbr{\rbr{\pg(\bn,z)-\bar\rho_g(z)}\rbr{\pg(\bn',z')-\bar\rho_g(z')}}=\bar\rho_g(z)\bar{\rho}_g(z')\xi_g(\bn\cdot\bn',z,z')\,, \label{eq:clust}
\eeq
where $\xi_g(\bn\cdot\bn',z,z')$ is the galaxy 2-point correlation function. Note that, due to statistical isotropy, $\xi_g(\bn\cdot\bn',z,z')$ depends only on the angle between $\bn$ and $\bn'$ and on the two redshifts $z$, $z'$. Our aim is to determine how the clustering of galaxies described by the correlation function $\xi_g(\bn\cdot\bn',z,z')$ impacts the HD correlation. In a given cosmological model, this correlation function can be easily calculated using Boltzmann codes such as \texttt{CLASS}~\cite{Lesgourgues:2011re}. Moreover, it has been measured by various surveys, such as WiggleZ~\cite{Blake_2011}, SDSS~\cite{Howlett:2014opa}, VIPERS~\cite{Pezzotta:2016gbo}, BOSS~\cite{Alam2016:1607.03155v1} and eBOSS~\cite{eBOSS:2020yzd}.  

\subsection{Ensemble average in one specific realization: conditional averaging} \label{sec:condition}

The case discussed in section~\ref{sec:noclust} describes a situation where the distribution of sources is homogeneous and isotropic. This corresponds to an average over all possible density realizations of the universe, which eliminates the directional dependence assuming that the universe is statistically isotropic, i.e.~does not have any preferred direction.\footnote{Indeed, the average on the left-hand side of Eq.~\eqref{eq:hh_noclust} is eliminating all stochasticity, hence the right-hand side is as well non-stochastic. The only way to obtain a direction-dependent spectral density after eliminating stochasticity is by breaking statistical isotropy, i.e.~by assuming that there would exist some special direction in our Universe.} In practice, anisotropies are typically not specifically mentioned, and the average over all realizations can be treated as an average done at a special realization, $\pg(\bn,z)=\bar\rho_g(z)$, where sources are isotropically distributed. In this spirit, Eq.~\eqref{eq:hh_noclust} can be written as
\beq
\left\langle h_A^\ast(f,\mathbf{n})h_B(f',\mathbf{n'})\,\big\vert\,\pg(\bn,z)=\bar\rho_g(z)\right\rangle =\frac 12 \delta_{AB}\,\delta(f-f')\bar S_h(f)\frac{\delta^2(\mathbf{n},\mathbf{n'})}{4\pi}\,.
\eeq
The expression on the left is a \textit{conditional ensemble average}: it represents the average over all realizations of $h_A$ (i.e.\ the expectation value) of the product $h_A^\ast(f,\bn)h_B(f',\mathbf{n'})$, 
assuming a \emph{fixed} realization of the density field $\pg(\bn,z)=\bar{\rho}_g(z)$. 
On the right-hand side, $\bar{S}_h(f)$ determines the variance (over all possible realizations of $h_A$) of the field $h_A$, in a universe where the galaxy density distribution is perfectly homogeneous. Since the density distribution does not depend on direction in this case, $\bar S_h(f)$ is also independent on $\bn$. Moreover, for a fixed realization of the density, $h_A$ is a Gaussian random field.

Let us now consider a different realization that is not perfectly homogeneous and isotropic: $\pg(\bn,z)=\rho_g(\bn,z)=\bar\rho_g(z)(1+\delta_g(\bn,z))$. Here, $\delta_g(\bn,z)$ is the galaxy over- or underdensity, which can thus be positive or negative. 
In such a realization, Eq.~\eqref{eq:hh_noclust} changes to
\beq
\left\langle h_A^\ast(f,\mathbf{n})h_B(f',\mathbf{n'})\,\big\vert \,\pg(\bn,z)=\rho_g(\bn,z)\right\rangle =\frac 12 \delta_{AB}\,\delta(f-f') S_h(f,\bn)\frac{\delta^2(\mathbf{n},\mathbf{n'})}{4\pi}\,. \label{eq:hh_clust}
\eeq
As before, $h_A(f, \mathbf{n})$ is a Gaussian random field given the specific realization $\rho_g(\bn,z)$, with the variance at each point determined by the spectral density $S_h(f,\bn)$ and no correlations between distinct angles, frequencies or polarization states. The only difference to before is that the spectral density $S_h(f,\bn)$ now depends on direction $\bn$, since the distribution of galaxies (and therefore the distribution of GW sources) is not isotropic in this realization of the universe. We can decompose the spectral density $S_h(f,\bn)$ into an isotropic part $\bar S_h(f)$, denoting the spectral density in an isotropic realization, and $\delta S_h(f,\mathbf{n})$, encoding the departure from isotropy in the realization $\rho_g(\bn, z)$,
\beq
S_h(f,\mathbf{n})=\bar S_h(f)+\delta S_h(f,\mathbf{n})=\bar S_h(f)\sbr{1+\int\mathrm d z\,b_{\rm GW}(f,z)\delta_g(\mathbf{n},z)}\,. \label{eq:sh_clust}
\eeq
 Here, $b_{\rm GW}(f,z)$ determines the relation between the galaxy overdensity $\delta_g(\bn,z)$ and the GW spectral density. 
Since we assume that the distribution of GW sources follows the distribution of galaxies, $b_{\rm GW}(f,z)$ does not depend on direction $\bn$. Its value, as well as its dependence on redshift and on frequency, is related to the astrophysical model describing the formation of GW sources and will need to be specified once numerical calculations are performed. In this work, to develop a theoretical framework broadly applicable to many astrophysical models, we keep $b_{\rm GW}(f,z)$ as a generic quantity.

\subsection{Total ensemble averages}

Since we aim to determine the variance of the HD correlation due to galaxy clustering, we need to calculate the ensemble average of $h_A$ over all realizations of the galaxy density field $\pg(\bn,z)$. For ease of notation, we first introduce a more concise notation for the ensemble average over $h$ in a fixed realization $\rho_g$,
\beq
\left\langle h_A^\ast(f,\mathbf{n})h_B(f',\mathbf{n'})\right\rangle_{h\vert\rho_g}\equiv\left\langle h_A^\ast(f,\mathbf{n})h_B(f',\mathbf{n'})\,\big\vert\, \pg(\bn,z)=\rho_g(\bn,z)\right\rangle \,.
\eeq
The full ensemble average, i.e.\ the average over all realizations of $h$ and all realizations of the density is then given by\footnote{Similar two-step averaging procedures have been applied in the context of anisotropies in the energy density of a stochastic GW background, see Refs.~\cite{Jenkins:2018lvb, Jenkins:2019nks, renzini2022}. Here, we apply this procedure, for the first time, to compute the impact of galaxy clustering on the variance of the HD correlation.}
\beq
\left\langle h_A^\ast(f,\mathbf{n})h_B(f',\mathbf{n'})\right\rangle=\angbr{\left\langle h_A^\ast(f,\mathbf{n})h_B(f',\mathbf{n'})\right\rangle_{h\vert\rho_g}}_{\pg}\,.
\eeq
 Here, we first compute the conditional ensemble average, and then integrate over the probability distribution of $\pg(\bn,z)$. This way, we obtain
\beq
\label{eq:hh_fullav}
\left\langle h_A^\ast(f,\mathbf{n})h_B(f',\mathbf{n'})\right\rangle=\angbr{\frac 12 \delta_{AB}\,\delta(f-f') S_h(f,\bn)\frac{\delta^2(\mathbf{n},\mathbf{n'})}{4\pi}}_{\pg}=\frac 12 \delta_{AB}\,\delta(f-f') 
\bar S_h(f)\frac{\delta^2(\mathbf{n},\mathbf{n'})}{4\pi}\,,
\eeq
where we used Eq.~\eqref{eq:sh_clust} and the fact that $\angbr{\delta_g(\bn,z)}_{\pg}=0$ (since $\angbr{\pg(\bn,z)}=\bar\rho_g(z)$). Hence, the variance of $h_A$ over all realizations of the density field is indeed the same as the variance of $h_A$ in a perfectly isotropic realization. This is not surprising since the mean of the density field over all realizations, $\bar\rho_g(z)$, is itself homogeneous and isotropic. 

With the tools developed in this section, we can now determine how the galaxy density field impacts the HD correlations and their variance. For clarity, we provide a summary of our notations for the different kinds of ensemble averages in Table~\ref{Table_Averages}.

\begin{table*}[!t]
  \centering
  \normalsize
  \renewcommand{\arraystretch}{1.5}
		\begin{tabular}[c]{ p{0.25 \textwidth}  p{0.75\textwidth}  }
				\hline \hline     
				\centering{$\langle\dots\rangle$} & Total ensemble average over all realizations of the universe, i.e.\ over all GW and density field realizations\\
                \centering{$\left\langle\dots\big\vert \pg(\bn,z)=\rho_g(\bn, z)\right\rangle$} & Conditional ensemble average over all GW realizations, for a fixed galaxy density realization \\
            \centering{$\left\langle\dots\right\rangle_{h\vert \rho_g}$} & A shorter notation for the one in the line above \\
                \centering{$\langle\dots\rangle_{\pg}$} & Ensemble average over all density field realizations of the universe \\
				\hline \hline
		\end{tabular}
  \caption{\label{table_summary}  We list our notations for the different kinds of ensemble averages appearing in this work.} 
   \label{Table_Averages}
\end{table*}

\section{Variance of the HD correlation due to galaxy clustering}
\label{Sec:HDvariance}

The HD correlation describes the correlation of timing residuals between pulsars, due to the passing of GWs. Following Ref.~\cite{Allen:2022dzg}, we calculate the redshift, $Z_a(t)$, induced by the GW on a pulse emitted by pulsar $a$ located in direction ${\bf n}_a$
and observed at time $t$ on Earth:
\be
Z_a(t)\equiv \sum_A \int {\rm d}f \int \dd \bn\, F_a^A(\bn)h_A(f, \bn) \E^{2\pi i f t}\left(1-\E^{-2\pi i 
 f \tau_a (1+\bn\cdot{\bf n}_a)}\right)\,.
\ee
Here, $\tau_a$ is the light travel time between emission and observation, and the antenna pattern functions $F_a^A(\bn)$ are defined as 
\be
 F_a^A(\bn)\equiv\frac{n_a^i n_a^j e_{ij}^A(\bn)}{2(1+\bn\cdot \bn_a)}\,.
\ee
Note that the property $h_A(-f,\bn)=h_A^\ast(f,\bn)$ implies that $Z_a(t)\in\mathbb{R}$. Since the redshift of a pulsar is related to its timing residual by time differentiation, the goal of PTA observations can be formulated as the search for correlations between the redshifts of pulsars.
We define therefore
$\rho_{ab}$ as the product of the redshifts of two pulsar signals averaged over the observation time $T$,\footnote{In this paper, $\rho_{ab}$ will always be used as defined in Eq.~\eqref{Eq:rho_ab}, and should not be confused with a density.} 
\begin{align}
    \rho_{ab}&\equiv\overline{Z_aZ_b} =\overline{Z_aZ_b^\ast} = \frac{1}{T}\int_{-T/2}^{T/2} {\mathrm d}t \  Z_aZ_b^\ast\nonumber\\
    &=\sum_{A,A'}\int\mathrm df\int\mathrm df'\int\mathrm d\bn\int\mathrm d\bn' R^A_a(f,\bn)^\ast R_b ^{A'}(f',\bn')h_A^\ast(f,\bn)h_{A'}(f',\bn')\mbox{sinc}\rbr{\pi(f-f')T}\,, \label{Eq:rho_ab}
\end{align}
where we followed Ref.~\cite{Allen:2022dzg} in the use of the shorthand notation,
\begin{equation}
    R^A_a(f,\bn)\equiv\rbr{1-\E^{-2\pi i f\tau_a(1+\bn\cdot\bn_a)}}F^A_a(\bn)\,.
\end{equation}
The function $\mbox{sinc}(x)\equiv \sin(x)/x$ in Eq.~\eqref{Eq:rho_ab} arises from the time averaging process:
\be
\frac{1}{T}\int_{-T/2}^{T/2}\dd t \ \E^{-2 \pi i f t}\E^{2 \pi i f' t} = \frac{\sin\left(\pi(f-f')T\right)}{\pi(f-f')T}\,.
\ee

In the standard case of isotropically distributed sources, the expectation value $\angbr{\rho_{ab}}$ over field realizations of $h_A$ leads to the standard HD angular correlation, as is done e.g.~in Ref.~\cite{Allen:2022dzg}. When sources are anisotropically distributed, however, the expectation value of $\rho_{ab}$ depends on the specific realization through the spectral density $S_h(f,\bn)$, as we see from Eq.~\eqref{eq:hh_clust}. We can compute the ensemble average of $\rho_{ab}$ over all realizations of the galaxy density, and using Eq.~\eqref{eq:hh_fullav}, we see immediately that we recover the standard HD curve predicted in Ref.~\cite{Hellings:1983fr}. This is not surprising, since the ensemble average of the galaxy density is itself homogeneous and isotropic.
For future reference, we write this property of $\rho_{ab}$ as 
\be
\angbr{\rho_{ab}} = \angbr{\rho_{ab}}^\text{st.}\,. \label{eq:HDremains}
\ee
In other words, even when taking anisotropies into account, $\angbr{\rho_{ab}}$ takes the ``standard" form $\angbr{\rho_{ab}}^\text{st.}$, namely the one that it has in computations assuming isotropy. The relevant question is to determine to which extent the observed HD correlation in our specific realization of the universe may differ from the expectation value. To assess this, we need to compute the variance of $\rho_{ab}$. Before proceeding with the calculation, we note that $\rho_{ab}$ is defined involving only one pair of pulsars. In practice, PTA surveys track many pulsars forming a large number of pairs, which needs to be taken into account when evaluating the variance for a realistic scenario~\cite{Allen:2022ksj, Allen:2022dzg, Romano:2023zhb}. Indeed, the concepts presented in the previous sections are valid independently of how many pulsars are considered. Here, we demonstrate how our formalism is applied in the case of a single pulsar pair. {In our companion paper, Ref.~\cite{Grimm:2024hgi}, we present results for the observational limits of a single as well as infinitely many pulsar pairs.}

From Eq.~\eqref{Eq:rho_ab} we have
\begin{align}
    &\rho^2_{ab}=\sum_{A, A', A'', A'''}\iiiint\mathrm df\mathrm df'\mathrm df''\mathrm df'''\iiiint\mathrm d\bn\,\mathrm d\bn'\mathrm d\bn''\mathrm d\bn'''\mbox{sinc}\rbr{\pi(f-f')T}\mbox{sinc}\rbr{\pi(f''-f''')T} \nonumber \\
    &\times R^{A}_a(f,\bn)^\ast R^{A'}_b(f',\bn') R^{A''}_a(f'',\bn'')R^{A'''}_b(f''',\bn''')^\ast h_A^\ast(f,\bn) h_{A'}(f',\bn') h_{A''}(f'',\bn'') h_{A'''}^\ast(f''',\bn''')\,. 
    \label{Eq:rho2}
\end{align}
The variance is given by
\beq
\angbr{\rho^2_{ab}}  - \angbr{\rho_{ab}}^2=\angbr{\angbr{\rho^2_{ab}}\cond}_{\pg}  - \angbr{\angbr{\rho_{ab}}\cond}^2_{\pg}\,.
\eeq
For the inner ensemble average we use Eq.~\eqref{eq:hh_clust}, and the fact that the field $h_A$ is Gaussian given any fixed realization of the density field, as discussed in section~\ref{sec:condition}. We can thus apply Isserlis' theorem (also known as Wick's theorem) to the conditioned field. This theorem implies that for a Gaussian field, correlations of four fields can be expressed as sums of products of correlations of two fields. Our calculation
 is following Appendix~C of Ref.~\cite{Allen:2022dzg} with the extension that now, we need to take the anisotropic term of $S_h(f,\bn)$ in Eq.~\eqref{eq:sh_clust} into account. More specifically, Isserlis' theorem reads
\begin{align}
    &\angbr{h_A^\ast(f,\bn) h_{A'}(f',\bn') h_{A''}(f'',\bn'') h_{A'''}^\ast(f''',\bn''')}{\cond} =  \nonumber \\
    &\angbr{h_A^\ast(f,\bn)h_{A'}(f',\bn')}{\cond}\angbr{h_{A''}^\ast(-f'',\bn'')h_{A'''}(-f''',\bn''')}{\cond} \nonumber \\
&+\angbr{h_A^\ast(f,\bn)h_{A''}(f'',\bn'')}{\cond} \angbr{h_{A'}^\ast(-f',\bn')h_{A'''}(-f''',\bn''')}{\cond} \nonumber \\
&+\angbr{h_A^\ast(f,\bn)h_{A'''}(-f''',\bn''')}{\cond}\angbr{h_{A'}^\ast(-f',\bn')h_{A''}(f'',\bn'')}{\cond}\,,
\end{align}
where negative signs in the arguments appear making use of the relation $h_A(f,\bn)=h_A^\ast(-f,\bn)$. Applying Eq.~\eqref{eq:hh_clust}, we find that
\begin{align}
&\angbr{h_A^\ast(f,\bn) h_{A'}(f',\bn') h_{A''}(f'',\bn'') h_{A'''}^\ast(f''',\bn''')}{\cond} = \nnn
&\frac 14\delta_{AA'}\delta_{A''A'''}\frac{\delta^2(\bn,\bn')}{4\pi}\frac{\delta^2(\bn'',\bn''')}{4\pi}\delta(f-f')\delta(f''-f''')S_h(f,\bn)S_h(f'',\bn'') \nnn
&+\frac 14\delta_{AA''}\delta_{A'A'''}\frac{\delta^2(\bn,\bn'')}{4\pi}\frac{\delta^2(\bn',\bn''')}{4\pi}\delta(f-f'')\delta(f'-f''')S_h(f,\bn)S_h(f',\bn') \nnn
&+\frac 14\delta_{AA'''}\delta_{A'A''}\frac{\delta^2(\bn,\bn''')}{4\pi}\frac{\delta^2(\bn',\bn'')}{4\pi}\delta(f+f''')\delta(f'+f'')S_h(f,\bn)S_h(f',\bn')\,.
\label{eq:4point}
\end{align}

Whenever products of the type $R^{A\ast}_a R^{A}_b$ are considered in integrals, it is common to use the approximation that $\tau_{a, b}f\gg1$ for situations of practical interest, and thus functions with unequal and highly oscillating phases average out:
\be
F^{A}_a F^{A}_b \ \left[1-\E^{2\pi i  f\tau_a(1+\bn\cdot \bn_a)}\right]\left[1-\E^{-2\pi i f\tau_b (1+\bn\cdot \bn_b)}\right] \simeq F^{A}_a F^{A}_b\ (1+\delta_{ab})\,.
\label{eq:fast_oscillations}
\ee
This approximation is equivalent to considering only the correlation of the Earth-term for two distinct pulsars. The contribution of the pulsar term is only relevant for the auto-correlation, i.e.~with identical pulsars $a=b$, and falls off fast for non-zero angular or radial separations (see e.g.\ Ref.~\cite{Mingarelli:2014xfa} or Appendix~C.2 of Ref.~\cite{Allen:2022dzg} for the contributions of pulsar terms). 
 Thus, this term can be neglected when $a\neq b$. The auto-correlation ($a=b$) term, on the other hand, carries contributions from the Earth and pulsar terms, and therefore the corresponding terms are twice as large as those for $a\neq b$. 

With this further approximation, substituting Eq.~\eqref{eq:4point} in Eq.~\eqref{Eq:rho2} leads to
\begin{align}
\angbr{\rho_{ab}^2}{\cond}=&\frac 14 \iint \mathrm df\,\mathrm df'\int \frac{\mathrm d\bn}{{4\pi}}\,S_h(f,\bn)\chi_{ab}(\bn)\int\frac{\mathrm d\bn'}{{4\pi}}\,S_h(f',\bn')\chi_{ab}(\bn') \nonumber \\
&+ \iint \mathrm df\,\mathrm df'\mathrm{sinc}^2\rbr{\pi(f-f')T}\int\frac{\mathrm d\bn}{{4\pi}}\,S_h(f,\bn)\chi_{aa}(\bn)\int\frac{\mathrm d\bn'}{{4\pi}}\,S_h(f',\bn')\chi_{bb}(\bn') \nnn
&+\frac 14 \iint \mathrm df\,\mathrm df'\mathrm{sinc}^2\rbr{\pi(f-f')T}\int\frac{\mathrm d\bn}{{4\pi}}\,S_h(f,\bn)\chi_{ab}(\bn)\int\frac{\mathrm d\bn'}{{4\pi}}\,S_h(f',\bn')\chi_{ab}(\bn')\,,
\label{eq:rhoabSh}
\end{align}
where
\beq
\chi_{ab}(\bn)=\sum_{A=+, \times} F_a^A(\bn)F_b^A(\bn)\,,
\eeq
and we focused on non-vanishing pulsar pair separations only, i.e.~$\rho_{ab}$ with $a\neq b$ (note that terms $\propto \chi_{aa},\chi_{bb}$ may nevertheless appear, for which we consider contributions from both the pulsar and Earth terms as discussed above). 

The GW spectral density, $S_h(f,\bn)$, appearing in Eq.~\eqref{eq:rhoabSh} can be expressed as the sum of an isotropic and anisotropic contribution, see Eq.~\eqref{eq:sh_clust}. We therefore split the expression for $\angbr{\rho^2_{ab}}{\cond}$ into a ``standard" and a ``clustering" part,
\begin{align}
\angbr{\rho_{ab}^2}{\cond}=\angbr{\rho_{ab}^2}^\st\cond+\angbr{\rho_{ab}^2}^\clust\cond\,.
\end{align}
The standard terms,
\begin{align}
\angbr{\rho_{ab}^2}^\st\cond=&\frac 14 \iint \mathrm df\,\mathrm df'\int\frac{\mathrm d\bn}{{4\pi}}\,\bar S_h(f)\chi_{ab}(\bn)\int\frac{\mathrm d\bn'}{{4\pi}}\,\bar S_h(f')\chi_{ab}(\bn') \nonumber \\
&+ \iint \mathrm df\,\mathrm df'\mathrm{sinc}^2\rbr{\pi(f-f')T}\int\frac{\mathrm d\bn}{{4\pi}}\,\bar S_h(f)\chi_{aa}(\bn)\int\frac{\mathrm d\bn'}{{4\pi}}\,\bar S_h(f')\chi_{bb}(\bn') \nnn
&+\frac 14 \iint \mathrm df\,\mathrm df'\mathrm{sinc}^2\rbr{\pi(f-f')T}\int \frac{\mathrm d\bn}{{4\pi}}\,\bar{S}_h(f)\chi_{ab}(\bn)\int\frac{\mathrm d\bn'}{{4\pi}}\,\bar{S}_h(f')\chi_{ab}(\bn')\,,
\label{eq:rhoab_avst}
\end{align}
are exactly those specified in Eq.~(C25) of Ref.~\cite{Allen:2022dzg}. The clustering terms, on the other hand, arise from the anisotropic part of $S_h(f,\bn)$ and are given by 
\begin{align}
{\angbr{\rho^2_{ab}}^\clust\cond}=&\iint\mathrm dz\,\mathrm dz'\left[G(z)G(z')+\Gamma(z,z')\right]\iint\frac{\mathrm d\bn}{4\pi}\frac{\mathrm d\bn'}{4\pi} \delta_g(\bn,z)\delta_g(\bn',z')\chi_{ab}(\bn)\chi_{ab}(\bn') \nonumber \\
&+4 \iint\mathrm dz\,\mathrm dz'\,\Gamma(z,z') \iint\frac{\mathrm d\bn}{4\pi}\frac{\mathrm d\bn'}{4\pi}\delta_g(\bn,z)\delta_g(\bn',z') \chi_{aa}(\bn)\chi_{bb}(\bn') \nnn
&+\mbox{terms linear in  $\delta_g(\bn,z)$}\,,
\label{eq:rhoab_avcl}
\end{align}
where
\begin{align}
G(z)&\equiv\int_0^\infty \mathrm df\,\bar S_h(f)b_{\rm GW}(f,z)\,, \nonumber \\
\Gamma(z,z')&\equiv\frac 14\iint\mathrm df\,\mathrm df'\mbox{sinc}^2\rbr{\pi(f-f')T}
\bar S_h(f)\bar S_h(f')b_{\rm GW}(f,z)b_{\rm GW}({f'},z')\,,
\end{align}
are measures of the strain weighted by the quantity $b_{\rm GW}(f,z)$ relating galaxy overdensity and GW intensity.\footnote{We note that, in the case where $b_{\rm GW}(f,z)=1$, $G$ and $\Gamma$ respectively correspond to the quantities $h^2$ and $\mathscr{h}^4$ defined in Ref.~\cite{Allen:2022dzg}.}
We note that, while terms linear in $\delta_g$  appear in  Eq.~\eqref{Eq:2ndmom_rho_ab_clustering}, they vanish in the full ensemble average since $\angbr{\delta_g(\bn,z)}_{\pg}=0$. 

Taking the ensemble average of Eqs.~\eqref{eq:rhoab_avst} and~\eqref{eq:rhoab_avcl} over all realizations of the galaxy density field leads to
\begin{align}
\angbr{\rho_{ab}^2}=\angbr{\angbr{\rho_{ab}^2}{\cond}}_{\pg}=\angbr{\rho_{ab}^2}^\st+\angbr{\rho_{ab}^2}^\clust\,,
\end{align}
where, for the standard terms which are independent of the density field, 
\begin{equation}    \angbr{\rho_{ab}^2}^\st=\angbr{\rho_{ab}^2}^\st\cond
\end{equation}
and, for the clustering terms,
\begin{align}
\angbr{\rho^2_{ab}}^\clust=&\iint\mathrm dz\,\mathrm dz'\left[G(z) G(z')+ \Gamma(z,z')\right]\iint\frac{\mathrm d\bn}{4\pi}\frac{\mathrm d\bn'}{4\pi} \xi_g(\bn\cdot\bn',z,z')\chi_{ab}(\bn)\chi_{ab}(\bn') \nonumber \\ 
&+4 \iint\mathrm dz\,\mathrm dz'\Gamma(z,z') \iint\frac{\mathrm d\bn}{4\pi}\frac{\mathrm d\bn'}{4\pi} \xi_g(\bn\cdot\bn',z,z')\chi_{aa}(\bn)\chi_{bb}(\bn')\,. \label{Eq:2ndmom_rho_ab_clustering}
\end{align}
Here, we have used Eq.~\eqref{eq:clust} in the form $\angbr{\delta_g(\bn,z)\delta_g(\bn',z')}=\xi_g(\bn\cdot\bn',z,z')$. Given Eq.~\eqref{eq:HDremains}, which states that $\angbr{\rho_{ab}}$ is unaffected by clustering, we have
\beq
\angbr{\rho^2_{ab}} - \angbr{\rho_{ab}}^2= \angbr{\rho^2_{ab}}^\text{st.} -\left(\angbr{\rho_{ab}}^\text{st.}\right)^2+\angbr{\rho^2_{ab}}^\clust\,.
\eeq
Eq.~\eqref{Eq:2ndmom_rho_ab_clustering} presents the main result of our work: an analytical expression encoding the effect of galaxy clustering (described by the galaxy correlation function $\xi_g$) on the HD variance.  In other words, this provides a measure of how much we expect the measured correlation of timing residuals to differ, due to galaxy clustering, from the idealized HD curve in a real measurement. 
We comment on the positivity of the contribution $\angbr{\rho^2_{ab}}^\clust$ to the total HD variance in Appendix~\ref{sec:Positivity}.

We close this section by noting that, instead of the galaxy correlation function, an alternative measure of galaxy clustering is provided by the angular power spectrum $C_l(z,z')$,
\begin{equation}
\xi_g(\bn\cdot\bn',z,z')=\frac{1}{4\pi}\sum_l (2l+1)C_l(z,z')\mathcal{P}_l(\bn\cdot\bn')\,,
\end{equation}
with $\mathcal{P}_l$ being the Legendre polynomials. This leads to
\begin{align}
\angbr{\rho^2_{ab}}^\clust= 
&\iint\mathrm dz\,\mathrm dz'\left[G(z)G(z')+\Gamma(z,z')\right]\sum_{l}\frac{2l+1}{4\pi}C_l(z,z')\iint\frac{\mathrm d\bn}{4\pi}\frac{\mathrm d\bn'}{4\pi}\,\mathcal{P}_l(\bn\cdot\bn')\chi_{ab}(\bn)\chi_{ab}(\bn') \nonumber \\
&+4 \iint\mathrm dz\,\mathrm dz'\,\Gamma(z,z') \sum_{l}\frac{2l+1}{4\pi}C_l(z,z')\iint\frac{\mathrm d\bn}{4\pi}\frac{\mathrm d\bn'}{4\pi}\mathcal{P}_l(\bn\cdot\bn') \chi_{aa}(\bn)\chi_{bb}(\bn')\,. \label{Eq:2ndmom_rho_ab_clustering_Cl}
\end{align}

As a further alternative, one can also choose to work with the galaxy power spectrum $P_g(k,z,z')$, 
\begin{equation}
\angbr{\delta_g(\mathbf{k},z)\delta_g^\ast(\mathbf{k}',z')}=(2\pi)^3\delta_D(\mathbf{k}-\mathbf{k'})P_g(k,z,z')\,.
\end{equation}
Here, $\delta_g(\mathbf{k},z)$ is the 3D Fourier transform of the galaxy overdensity $\delta_g(\bn,z)$.
 Using the galaxy power spectrum, Eq.~\eqref{Eq:2ndmom_rho_ab_clustering} can be simplified and expressed as one instead of two integrals over redshift by using the Limber approximation~\cite{limber1953analysis}, a commonly used approximation in large-scale structure observables. In particular, following Ref.~\cite{Bartelmann:1999yn}, we obtain 
\begin{align}
\angbr{\rho^2_{ab}}^\clust=&\int\mathrm dz\left[{G^2(z)}+\Gamma(z,z)\right]{H(z)}\int\frac{k\,\mathrm dk}{2\pi}P_g(k,z)\iint\frac{\mathrm d\bn}{4\pi}\frac{\mathrm d\bn'}{4\pi}\chi_{ab}(\bn)\chi_{ab}(\bn') J_0\big(k r_z\,{\vert \bn -\bn'\vert }\big)\nonumber \\ 
&+4 \int\mathrm dz\,\Gamma(z,z){H(z)} \int\frac{k\,\mathrm dk}{2\pi}P_g(k,z) \iint\frac{\mathrm d\bn}{{4\pi}}\frac{\mathrm d\bn'}{{4\pi}} \chi_{aa}(\bn)\chi_{bb}(\bn')J_0\big(k r_z\,{\vert\bn-\bn'\vert}\big)\,, \label{Eq:2ndmom_Limber}
\end{align}
with $J_0$ denoting the 0th order Bessel function and $H(z)$ being the Hubble parameter.\footnote{The factor $H(z)$ appears due to the relation $\mathrm dz = H(z)\mathrm dr$, i.e.~the change of integration variable from redshift $z$ to comoving distance $r$, which is needed to apply the Limber approximation (cf.~Eq.~(2.83) in Ref.~\cite{Bartelmann:1999yn}).}

\section{Conclusion} \label{sec:conclusion}
This work introduces a rigorous theoretical framework to assess how the HD correlation, describing pulsar pair correlations caused by a stochastic GW background of astrophysical origin, is affected by anisotropies in the distribution of sources due to large-scale galaxy clustering. The presence of such anisotropies is naturally expected, since the GW sources (e.g.\ supermassive black holes at the center of galaxies) are expected to follow the distribution of matter in the Universe, which is itself anisotropically distributed.

We showed that to properly account for such anisotropies, it is necessary to consider not only ensemble averages over realizations of GW sources, but also ensemble averages over realizations of the galaxy density field. 
Modeling fluctuations around an isotropic universe by Gaussian random fields and performing ensemble averaging taking  two layers of stochasticity into account, we find that the expectation value of the HD correlation is unchanged with respect to that of an isotropic universe. 
However, we find that anisotropies lead to a new contribution to the variance of the correlation, which has not been considered in previous work~\cite{Allen:2022dzg, Allen:2022ksj, Romano:2023zhb, Allen:2024rqk}. Thereby, large-scale galaxy clustering impacts how much one expects observed timing residual correlations to deviate from the idealized HD curve.

This contribution is important to assess in view of future observations.  While the present-day PTA data has error bars sufficiently large to fully overlap with the HD curve \cite{NANOGrav:2023gor}, the precision of observations is expected to improve~\cite{babak2024forecasting}. Correctly modeling the variance of the HD correlation is therefore important in view of future observations. Indeed, a GW background of cosmological origin is only marginally affected by clustering (i.e.~only via relativistic line-of-sight effects, which give very small contributions that we neglect in this work). As such, the clustering induced contribution to the variance evaluated in this work, which is only present in the case of an astrophysical origin, is a new interesting observable which may provide us with precious information on the nature of the stochastic GW background. Moreover, since various work on PTA observations has investigated the possibility to search for new physics (see e.g.\ Refs.~\cite{liang2023test,NANOGrav:2023hvm,omiya2023hellings}), it is crucial to properly evaluate timing residual correlations accounting for all fundamental contributions to their variance, to enable an accurate comparison between future high-precision data and models.  

In a companion paper~\cite{Grimm:2024hgi}, we apply our theoretical formalism to specific population models of supermassive black holes, considering as well the scenario of many pulsar pairs. In particular, in order to precisely quantify the size of the clustering induced variance, we need to determine how the distribution of sources is linked to the distribution of galaxies. This depends closely on the details of the source formation and evolution. We expect that different formation scenarios, leading to different source distributions, may result in different magnitudes of the HD variance, thus influencing how much we expect observations of the HD correlation to differ from the idealized HD curve. This work, in combination with its companion paper~\cite{Grimm:2024hgi}, provides a new framework to study the link between two important observational probes of the properties of the Universe: the large-scale clustering of galaxies and the stochastic GW background observed via PTAs. 

\section*{Acknowledgements}
We acknowledge discussions with Bruce Allen. N.G. and C.B. acknowledge funding from the European Research Council (ERC) under the European Union’s Horizon 2020 research and innovation program (Grant agreement No.~863929; project title ``Testing the law of gravity with novel large-scale structure observables"). C.B. is also supported by the Swiss National Science Foundation. The work of G.C. is supported by CNRS and G.C. and M.P. acknowledge support from the Swiss National Science Foundation (Ambizione grant, ``Gravitational wave propagation in the clustered universe").

\bibliographystyle{Bonvinetal}
\bibliography{PTASpectrumRefs}

\ifx\mcitethebibliography\mciteundefinedmacro
\PackageError{LHCb.bst}{mciteplus.sty has not been loaded}
{This bibstyle requires the use of the mciteplus package.}\fi
\providecommand{\href}[2]{#2}
\begin{mcitethebibliography}{10}
\mciteSetBstSublistMode{n}
\mciteSetBstMaxWidthForm{subitem}{\alph{mcitesubitemcount})}
\mciteSetBstSublistLabelBeginEnd{\mcitemaxwidthsubitemform\space}
{\relax}{\relax}

\bibitem{NANOGrav:2023gor}
NANOGrav, G.~Agazie {\em et~al.}, \ifthenelse{\boolean{articletitles}}{\emph{{The NANOGrav 15 yr Data Set: Evidence for a Gravitational-wave Background}}, }{}\href{https://doi.org/10.3847/2041-8213/acdac6}{Astrophys.\ J.\ Lett.\  \textbf{951} (2023) L8}, \href{http://arxiv.org/abs/2306.16213}{{\normalfont\ttfamily arXiv:2306.16213}}\relax
\mciteBstWouldAddEndPuncttrue
\mciteSetBstMidEndSepPunct{\mcitedefaultmidpunct}
{\mcitedefaultendpunct}{\mcitedefaultseppunct}\relax
\EndOfBibitem
\bibitem{Reardon:2023gzh}
D.~J. Reardon {\em et~al.}, \ifthenelse{\boolean{articletitles}}{\emph{{Search for an Isotropic Gravitational-wave Background with the Parkes Pulsar Timing Array}}, }{}\href{https://doi.org/10.3847/2041-8213/acdd02}{Astrophys.\ J.\ Lett.\  \textbf{951} (2023) L6}, \href{http://arxiv.org/abs/2306.16215}{{\normalfont\ttfamily arXiv:2306.16215}}\relax
\mciteBstWouldAddEndPuncttrue
\mciteSetBstMidEndSepPunct{\mcitedefaultmidpunct}
{\mcitedefaultendpunct}{\mcitedefaultseppunct}\relax
\EndOfBibitem
\bibitem{EPTA:2023sfo}
EPTA, J.~Antoniadis {\em et~al.}, \ifthenelse{\boolean{articletitles}}{\emph{{The second data release from the European Pulsar Timing Array I. The dataset and timing analysis}}, }{}\href{https://doi.org/10.1051/0004-6361/202346841}{Astron.\ Astrophys.\  \textbf{678} (2023) A48}, \href{http://arxiv.org/abs/2306.16224}{{\normalfont\ttfamily arXiv:2306.16224}}\relax
\mciteBstWouldAddEndPuncttrue
\mciteSetBstMidEndSepPunct{\mcitedefaultmidpunct}
{\mcitedefaultendpunct}{\mcitedefaultseppunct}\relax
\EndOfBibitem
\bibitem{Xu:2023wog}
H.~Xu {\em et~al.}, \ifthenelse{\boolean{articletitles}}{\emph{{Searching for the Nano-Hertz Stochastic Gravitational Wave Background with the Chinese Pulsar Timing Array Data Release I}}, }{}\href{https://doi.org/10.1088/1674-4527/acdfa5}{Res.\ Astron.\ Astrophys.\  \textbf{23} (2023) 075024}, \href{http://arxiv.org/abs/2306.16216}{{\normalfont\ttfamily arXiv:2306.16216}}\relax
\mciteBstWouldAddEndPuncttrue
\mciteSetBstMidEndSepPunct{\mcitedefaultmidpunct}
{\mcitedefaultendpunct}{\mcitedefaultseppunct}\relax
\EndOfBibitem
\bibitem{Shannon:2015ect}
R.~M. Shannon {\em et~al.}, \ifthenelse{\boolean{articletitles}}{\emph{{Gravitational waves from binary supermassive black holes missing in pulsar observations}}, }{}\href{https://doi.org/10.1126/science.aab1910}{Science \textbf{349} (2015) 1522}, \href{http://arxiv.org/abs/1509.07320}{{\normalfont\ttfamily arXiv:1509.07320}}\relax
\mciteBstWouldAddEndPuncttrue
\mciteSetBstMidEndSepPunct{\mcitedefaultmidpunct}
{\mcitedefaultendpunct}{\mcitedefaultseppunct}\relax
\EndOfBibitem
\bibitem{Chen:2019xse}
Z.-C. Chen, C.~Yuan, and Q.-G. Huang, \ifthenelse{\boolean{articletitles}}{\emph{{Pulsar Timing Array Constraints on Primordial Black Holes with NANOGrav 11-Year Dataset}}, }{}\href{https://doi.org/10.1103/PhysRevLett.124.251101}{Phys.\ Rev.\ Lett.\  \textbf{124} (2020) 251101}, \href{http://arxiv.org/abs/1910.12239}{{\normalfont\ttfamily arXiv:1910.12239}}\relax
\mciteBstWouldAddEndPuncttrue
\mciteSetBstMidEndSepPunct{\mcitedefaultmidpunct}
{\mcitedefaultendpunct}{\mcitedefaultseppunct}\relax
\EndOfBibitem
\bibitem{NANOGrav:2021flc}
NANOGrav, Z.~Arzoumanian {\em et~al.}, \ifthenelse{\boolean{articletitles}}{\emph{{Searching for Gravitational Waves from Cosmological Phase Transitions with the NANOGrav 12.5-Year Dataset}}, }{}\href{https://doi.org/10.1103/PhysRevLett.127.251302}{Phys.\ Rev.\ Lett.\  \textbf{127} (2021) 251302}, \href{http://arxiv.org/abs/2104.13930}{{\normalfont\ttfamily arXiv:2104.13930}}\relax
\mciteBstWouldAddEndPuncttrue
\mciteSetBstMidEndSepPunct{\mcitedefaultmidpunct}
{\mcitedefaultendpunct}{\mcitedefaultseppunct}\relax
\EndOfBibitem
\bibitem{NANOGrav:2023hvm}
NANOGrav, A.~Afzal {\em et~al.}, \ifthenelse{\boolean{articletitles}}{\emph{{The NANOGrav 15 yr Data Set: Search for Signals from New Physics}}, }{}\href{https://doi.org/10.3847/2041-8213/acdc91}{Astrophys.\ J.\ Lett.\  \textbf{951} (2023) L11}, \href{http://arxiv.org/abs/2306.16219}{{\normalfont\ttfamily arXiv:2306.16219}}\relax
\mciteBstWouldAddEndPuncttrue
\mciteSetBstMidEndSepPunct{\mcitedefaultmidpunct}
{\mcitedefaultendpunct}{\mcitedefaultseppunct}\relax
\EndOfBibitem
\bibitem{Vagnozzi:2023lwo}
S.~Vagnozzi, \ifthenelse{\boolean{articletitles}}{\emph{{Inflationary interpretation of the stochastic gravitational wave background signal detected by pulsar timing array experiments}}, }{}\href{https://doi.org/10.1016/j.jheap.2023.07.001}{JHEAp \textbf{39} (2023) 81}, \href{http://arxiv.org/abs/2306.16912}{{\normalfont\ttfamily arXiv:2306.16912}}\relax
\mciteBstWouldAddEndPuncttrue
\mciteSetBstMidEndSepPunct{\mcitedefaultmidpunct}
{\mcitedefaultendpunct}{\mcitedefaultseppunct}\relax
\EndOfBibitem
\bibitem{Blake_2011}
C.~Blake, S.~Brough, M.~Colless, C.~Contreras, W.~Couch {\em et~al.}, \ifthenelse{\boolean{articletitles}}{\emph{The {WiggleZ} dark energy survey: the growth rate of cosmic structure since redshift z=0.9}, }{}\href{https://doi.org/10.1111/j.1365-2966.2011.18903.x}{\mnras \textbf{415} (2011) 2876}, \href{http://arxiv.org/abs/1104.2948}{{\normalfont\ttfamily arXiv:1104.2948}}\relax
\mciteBstWouldAddEndPuncttrue
\mciteSetBstMidEndSepPunct{\mcitedefaultmidpunct}
{\mcitedefaultendpunct}{\mcitedefaultseppunct}\relax
\EndOfBibitem
\bibitem{Howlett:2014opa}
C.~Howlett, A.~Ross, L.~Samushia, W.~Percival, and M.~Manera, \ifthenelse{\boolean{articletitles}}{\emph{{The clustering of the {SDSS} main galaxy sample \textendash{} II. Mock galaxy catalogues and a measurement of the growth of structure from redshift space distortions at $z = 0.15$}}, }{}\href{https://doi.org/10.1093/mnras/stu2693}{\mnras \textbf{449} (2015) 848}, \href{http://arxiv.org/abs/1409.3238}{{\normalfont\ttfamily arXiv:1409.3238}}\relax
\mciteBstWouldAddEndPuncttrue
\mciteSetBstMidEndSepPunct{\mcitedefaultmidpunct}
{\mcitedefaultendpunct}{\mcitedefaultseppunct}\relax
\EndOfBibitem
\bibitem{Pezzotta:2016gbo}
A.~Pezzotta, S.~de~la Torre, J.~Bel, B.~R. Granett, L.~Guzzo {\em et~al.}, \ifthenelse{\boolean{articletitles}}{\emph{{The VIMOS Public Extragalactic Redshift Survey (VIPERS): The growth of structure at $0.5 < z < 1.2$ from redshift-space distortions in the clustering of the PDR-2 final sample}}, }{}\href{https://doi.org/10.1051/0004-6361/201630295}{\aap \textbf{604} (2017) A33}, \href{http://arxiv.org/abs/1612.05645}{{\normalfont\ttfamily arXiv:1612.05645}}\relax
\mciteBstWouldAddEndPuncttrue
\mciteSetBstMidEndSepPunct{\mcitedefaultmidpunct}
{\mcitedefaultendpunct}{\mcitedefaultseppunct}\relax
\EndOfBibitem
\bibitem{Alam2016:1607.03155v1}
S.~Alam, M.~Ata, S.~Bailey, F.~Beutler, D.~Bizyaev {\em et~al.}, \ifthenelse{\boolean{articletitles}}{\emph{The clustering of galaxies in the completed {SDSS-III} baryon oscillation spectroscopic survey: cosmological analysis of the {DR12} galaxy sample}, }{}\href{https://doi.org/10.1093/mnras/stx721}{\mnras \textbf{470} (2017) 2617}, \href{http://arxiv.org/abs/1607.03155}{{\normalfont\ttfamily arXiv:1607.03155}}\relax
\mciteBstWouldAddEndPuncttrue
\mciteSetBstMidEndSepPunct{\mcitedefaultmidpunct}
{\mcitedefaultendpunct}{\mcitedefaultseppunct}\relax
\EndOfBibitem
\bibitem{eBOSS:2020yzd}
eBOSS collaboration, S.~Alam, M.~Aubert, S.~Avila, C.~Balland, J.~E. Bautista {\em et~al.}, \ifthenelse{\boolean{articletitles}}{\emph{{Completed {SDSS-IV} extended Baryon Oscillation Spectroscopic Survey: Cosmological implications from two decades of spectroscopic surveys at the Apache Point Observatory}}, }{}\href{https://doi.org/10.1103/PhysRevD.103.083533}{\prd \textbf{103} (2021) 083533}, \href{http://arxiv.org/abs/2007.08991}{{\normalfont\ttfamily arXiv:2007.08991}}\relax
\mciteBstWouldAddEndPuncttrue
\mciteSetBstMidEndSepPunct{\mcitedefaultmidpunct}
{\mcitedefaultendpunct}{\mcitedefaultseppunct}\relax
\EndOfBibitem
\bibitem{Cusin:2017fwz}
G.~Cusin, C.~Pitrou, and J.-P. Uzan, \ifthenelse{\boolean{articletitles}}{\emph{{Anisotropy of the astrophysical gravitational wave background: Analytic expression of the angular power spectrum and correlation with cosmological observations}}, }{}\href{https://doi.org/10.1103/PhysRevD.96.103019}{Phys.\ Rev.\  \textbf{D96} (2017) 103019}, \href{http://arxiv.org/abs/1704.06184}{{\normalfont\ttfamily arXiv:1704.06184}}\relax
\mciteBstWouldAddEndPuncttrue
\mciteSetBstMidEndSepPunct{\mcitedefaultmidpunct}
{\mcitedefaultendpunct}{\mcitedefaultseppunct}\relax
\EndOfBibitem
\bibitem{Cusin:2017mjm}
G.~Cusin, C.~Pitrou, and J.-P. Uzan, \ifthenelse{\boolean{articletitles}}{\emph{{The signal of the gravitational wave background and the angular correlation of its energy density}}, }{}\href{https://doi.org/10.1103/PhysRevD.97.123527}{Phys.\ Rev.\  \textbf{D97} (2018) 123527}, \href{http://arxiv.org/abs/1711.11345}{{\normalfont\ttfamily arXiv:1711.11345}}\relax
\mciteBstWouldAddEndPuncttrue
\mciteSetBstMidEndSepPunct{\mcitedefaultmidpunct}
{\mcitedefaultendpunct}{\mcitedefaultseppunct}\relax
\EndOfBibitem
\bibitem{Cusin:2018rsq}
G.~Cusin, I.~Dvorkin, C.~Pitrou, and J.-P. Uzan, \ifthenelse{\boolean{articletitles}}{\emph{{First predictions of the angular power spectrum of the astrophysical gravitational wave background}}, }{}\href{https://doi.org/10.1103/PhysRevLett.120.231101}{Phys.\ Rev.\ Lett.\  \textbf{120} (2018) 231101}, \href{http://arxiv.org/abs/1803.03236}{{\normalfont\ttfamily arXiv:1803.03236}}\relax
\mciteBstWouldAddEndPuncttrue
\mciteSetBstMidEndSepPunct{\mcitedefaultmidpunct}
{\mcitedefaultendpunct}{\mcitedefaultseppunct}\relax
\EndOfBibitem
\bibitem{Pitrou:2019rjz}
C.~Pitrou, G.~Cusin, and J.-P. Uzan, \ifthenelse{\boolean{articletitles}}{\emph{{Unified view of anisotropies in the astrophysical gravitational-wave background}}, }{}\href{https://doi.org/10.1103/PhysRevD.101.081301}{Phys.\ Rev.\ D \textbf{101} (2020) 081301}, \href{http://arxiv.org/abs/1910.04645}{{\normalfont\ttfamily arXiv:1910.04645}}\relax
\mciteBstWouldAddEndPuncttrue
\mciteSetBstMidEndSepPunct{\mcitedefaultmidpunct}
{\mcitedefaultendpunct}{\mcitedefaultseppunct}\relax
\EndOfBibitem
\bibitem{Hellings:1983fr}
R.~w. Hellings and G.~s. Downs, \ifthenelse{\boolean{articletitles}}{\emph{{UPPER LIMITS ON THE ISOTROPIC GRAVITATIONAL RADIATION BACKGROUND FROM PULSAR TIMING ANALYSIS}}, }{}\href{https://doi.org/10.1086/183954}{Astrophys.\ J.\ Lett.\  \textbf{265} (1983) L39}\relax
\mciteBstWouldAddEndPuncttrue
\mciteSetBstMidEndSepPunct{\mcitedefaultmidpunct}
{\mcitedefaultendpunct}{\mcitedefaultseppunct}\relax
\EndOfBibitem
\bibitem{Allen:2022dzg}
B.~Allen, \ifthenelse{\boolean{articletitles}}{\emph{{Variance of the Hellings-Downs correlation}}, }{}\href{https://doi.org/10.1103/PhysRevD.107.043018}{Phys.\ Rev.\ D \textbf{107} (2023) 043018}, \href{http://arxiv.org/abs/2205.05637}{{\normalfont\ttfamily arXiv:2205.05637}}\relax
\mciteBstWouldAddEndPuncttrue
\mciteSetBstMidEndSepPunct{\mcitedefaultmidpunct}
{\mcitedefaultendpunct}{\mcitedefaultseppunct}\relax
\EndOfBibitem
\bibitem{Allen:2022ksj}
B.~Allen and J.~D. Romano, \ifthenelse{\boolean{articletitles}}{\emph{{Hellings and Downs correlation of an arbitrary set of pulsars}}, }{}\href{https://doi.org/10.1103/PhysRevD.108.043026}{Phys.\ Rev.\ D \textbf{108} (2023) 043026}, \href{http://arxiv.org/abs/2208.07230}{{\normalfont\ttfamily arXiv:2208.07230}}\relax
\mciteBstWouldAddEndPuncttrue
\mciteSetBstMidEndSepPunct{\mcitedefaultmidpunct}
{\mcitedefaultendpunct}{\mcitedefaultseppunct}\relax
\EndOfBibitem
\bibitem{Romano:2023zhb}
J.~D. Romano and B.~Allen, \ifthenelse{\boolean{articletitles}}{\emph{{Answers to frequently asked questions about the pulsar timing array Hellings and Downs curve}}, }{}\href{http://arxiv.org/abs/2308.05847}{{\normalfont\ttfamily arXiv:2308.05847}}\relax
\mciteBstWouldAddEndPuncttrue
\mciteSetBstMidEndSepPunct{\mcitedefaultmidpunct}
{\mcitedefaultendpunct}{\mcitedefaultseppunct}\relax
\EndOfBibitem
\bibitem{Allen:2024rqk}
B.~Allen and S.~Valtolina, \ifthenelse{\boolean{articletitles}}{\emph{{Pulsar Timing Array source ensembles}}, }{}\href{http://arxiv.org/abs/2401.14329}{{\normalfont\ttfamily arXiv:2401.14329}}\relax
\mciteBstWouldAddEndPuncttrue
\mciteSetBstMidEndSepPunct{\mcitedefaultmidpunct}
{\mcitedefaultendpunct}{\mcitedefaultseppunct}\relax
\EndOfBibitem
\bibitem{Bernardo:2022xzl}
R.~C. Bernardo and K.-W. Ng, \ifthenelse{\boolean{articletitles}}{\emph{{Pulsar and cosmic variances of pulsar timing-array correlation measurements of the stochastic gravitational wave background}}, }{}\href{https://doi.org/10.1088/1475-7516/2022/11/046}{JCAP \textbf{11} (2022) 046}, \href{http://arxiv.org/abs/2209.14834}{{\normalfont\ttfamily arXiv:2209.14834}}\relax
\mciteBstWouldAddEndPuncttrue
\mciteSetBstMidEndSepPunct{\mcitedefaultmidpunct}
{\mcitedefaultendpunct}{\mcitedefaultseppunct}\relax
\EndOfBibitem
\bibitem{Allen:2022bjz}
B.~Allen, \ifthenelse{\boolean{articletitles}}{\emph{{Will pulsar timing arrays observe the hellings and downs correlation curve?}}, }{}Frascati Phys.\ Ser.\  \textbf{74} (2022) 65\relax
\mciteBstWouldAddEndPuncttrue
\mciteSetBstMidEndSepPunct{\mcitedefaultmidpunct}
{\mcitedefaultendpunct}{\mcitedefaultseppunct}\relax
\EndOfBibitem
\bibitem{Allen:1996gp}
B.~Allen and A.~C. Ottewill, \ifthenelse{\boolean{articletitles}}{\emph{{Detection of anisotropies in the gravitational wave stochastic background}}, }{}\href{https://doi.org/10.1103/PhysRevD.56.545}{Phys.\ Rev.\ D \textbf{56} (1997) 545}, \href{http://arxiv.org/abs/gr-qc/9607068}{{\normalfont\ttfamily arXiv:gr-qc/9607068}}\relax
\mciteBstWouldAddEndPuncttrue
\mciteSetBstMidEndSepPunct{\mcitedefaultmidpunct}
{\mcitedefaultendpunct}{\mcitedefaultseppunct}\relax
\EndOfBibitem
\bibitem{Cusin:2018avf}
G.~Cusin, R.~Durrer, and P.~G. Ferreira, \ifthenelse{\boolean{articletitles}}{\emph{{Polarization of a stochastic gravitational wave background through diffusion by massive structures}}, }{}\href{https://doi.org/10.1103/PhysRevD.99.023534}{Phys.\ Rev.\ D \textbf{99} (2019) 023534}, \href{http://arxiv.org/abs/1807.10620}{{\normalfont\ttfamily arXiv:1807.10620}}\relax
\mciteBstWouldAddEndPuncttrue
\mciteSetBstMidEndSepPunct{\mcitedefaultmidpunct}
{\mcitedefaultendpunct}{\mcitedefaultseppunct}\relax
\EndOfBibitem
\bibitem{Cusin:2019jpv}
G.~Cusin, I.~Dvorkin, C.~Pitrou, and J.-P. Uzan, \ifthenelse{\boolean{articletitles}}{\emph{{Properties of the stochastic astrophysical gravitational wave background: astrophysical sources dependencies}}, }{}\href{https://doi.org/10.1103/PhysRevD.100.063004}{Phys.\ Rev.\ D \textbf{100} (2019) 063004}, \href{http://arxiv.org/abs/1904.07797}{{\normalfont\ttfamily arXiv:1904.07797}}\relax
\mciteBstWouldAddEndPuncttrue
\mciteSetBstMidEndSepPunct{\mcitedefaultmidpunct}
{\mcitedefaultendpunct}{\mcitedefaultseppunct}\relax
\EndOfBibitem
\bibitem{Cusin:2019jhg}
G.~Cusin, I.~Dvorkin, C.~Pitrou, and J.-P. Uzan, \ifthenelse{\boolean{articletitles}}{\emph{{Stochastic gravitational wave background anisotropies in the mHz band: astrophysical dependencies}}, }{}\href{https://doi.org/10.1093/mnrasl/slz182}{Mon.\ Not.\ Roy.\ Astron.\ Soc.\  \textbf{493} (2020) L1}, \href{http://arxiv.org/abs/1904.07757}{{\normalfont\ttfamily arXiv:1904.07757}}\relax
\mciteBstWouldAddEndPuncttrue
\mciteSetBstMidEndSepPunct{\mcitedefaultmidpunct}
{\mcitedefaultendpunct}{\mcitedefaultseppunct}\relax
\EndOfBibitem
\bibitem{Jenkins:2019nks}
A.~C. Jenkins, J.~D. Romano, and M.~Sakellariadou, \ifthenelse{\boolean{articletitles}}{\emph{Estimating the angular power spectrum of the gravitational-wave background in the presence of shot noise}, }{}\href{https://doi.org/10.1103/PhysRevD.100.083501}{Phys.\ Rev.\ D \textbf{100} (2019) 083501}\relax
\mciteBstWouldAddEndPuncttrue
\mciteSetBstMidEndSepPunct{\mcitedefaultmidpunct}
{\mcitedefaultendpunct}{\mcitedefaultseppunct}\relax
\EndOfBibitem
\bibitem{Alonso:2020mva}
D.~Alonso, G.~Cusin, P.~G. Ferreira, and C.~Pitrou, \ifthenelse{\boolean{articletitles}}{\emph{{Detecting the anisotropic astrophysical gravitational wave background in the presence of shot noise through cross-correlations}}, }{}\href{https://doi.org/10.1103/PhysRevD.102.023002}{Phys.\ Rev.\ D \textbf{102} (2020) 023002}, \href{http://arxiv.org/abs/2002.02888}{{\normalfont\ttfamily arXiv:2002.02888}}\relax
\mciteBstWouldAddEndPuncttrue
\mciteSetBstMidEndSepPunct{\mcitedefaultmidpunct}
{\mcitedefaultendpunct}{\mcitedefaultseppunct}\relax
\EndOfBibitem
\bibitem{renzini2022}
A.~I. Renzini, J.~D. Romano, C.~R. Contaldi, and N.~J. Cornish, \ifthenelse{\boolean{articletitles}}{\emph{Comparison of maximum-likelihood mapping methods for gravitational-wave backgrounds}, }{}\href{https://doi.org/10.1103/PhysRevD.105.023519}{Phys.\ Rev.\ D \textbf{105} (2022) 023519}\relax
\mciteBstWouldAddEndPuncttrue
\mciteSetBstMidEndSepPunct{\mcitedefaultmidpunct}
{\mcitedefaultendpunct}{\mcitedefaultseppunct}\relax
\EndOfBibitem
\bibitem{Mingarelli:2013dsa}
C.~M.~F. Mingarelli, T.~Sidery, I.~Mandel, and A.~Vecchio, \ifthenelse{\boolean{articletitles}}{\emph{{Characterizing gravitational wave stochastic background anisotropy with pulsar timing arrays}}, }{}\href{https://doi.org/10.1103/PhysRevD.88.062005}{Phys.\ Rev.\ D \textbf{88} (2013) 062005}, \href{http://arxiv.org/abs/1306.5394}{{\normalfont\ttfamily arXiv:1306.5394}}\relax
\mciteBstWouldAddEndPuncttrue
\mciteSetBstMidEndSepPunct{\mcitedefaultmidpunct}
{\mcitedefaultendpunct}{\mcitedefaultseppunct}\relax
\EndOfBibitem
\bibitem{Taylor:2013esa}
S.~R. Taylor and J.~R. Gair, \ifthenelse{\boolean{articletitles}}{\emph{{Searching For Anisotropic Gravitational-wave Backgrounds Using Pulsar Timing Arrays}}, }{}\href{https://doi.org/10.1103/PhysRevD.88.084001}{Phys.\ Rev.\ D \textbf{88} (2013) 084001}, \href{http://arxiv.org/abs/1306.5395}{{\normalfont\ttfamily arXiv:1306.5395}}\relax
\mciteBstWouldAddEndPuncttrue
\mciteSetBstMidEndSepPunct{\mcitedefaultmidpunct}
{\mcitedefaultendpunct}{\mcitedefaultseppunct}\relax
\EndOfBibitem
\bibitem{Gair:2014rwa}
J.~Gair, J.~D. Romano, S.~Taylor, and C.~M.~F. Mingarelli, \ifthenelse{\boolean{articletitles}}{\emph{{Mapping gravitational-wave backgrounds using methods from CMB analysis: Application to pulsar timing arrays}}, }{}\href{https://doi.org/10.1103/PhysRevD.90.082001}{Phys.\ Rev.\  \textbf{D90} (2014) 082001}, \href{http://arxiv.org/abs/1406.4664}{{\normalfont\ttfamily arXiv:1406.4664}}\relax
\mciteBstWouldAddEndPuncttrue
\mciteSetBstMidEndSepPunct{\mcitedefaultmidpunct}
{\mcitedefaultendpunct}{\mcitedefaultseppunct}\relax
\EndOfBibitem
\bibitem{Ali-Haimoud:2016mbv}
Y.~Ali-Haïmoud and M.~Kamionkowski, \ifthenelse{\boolean{articletitles}}{\emph{{Cosmic microwave background limits on accreting primordial black holes}}, }{}\href{https://doi.org/10.1103/PhysRevD.95.043534}{Phys.\ Rev.\  \textbf{D95} (2017) 043534}, \href{http://arxiv.org/abs/1612.05644}{{\normalfont\ttfamily arXiv:1612.05644}}\relax
\mciteBstWouldAddEndPuncttrue
\mciteSetBstMidEndSepPunct{\mcitedefaultmidpunct}
{\mcitedefaultendpunct}{\mcitedefaultseppunct}\relax
\EndOfBibitem
\bibitem{Grimm:2024hgi}
N.~Grimm, M.~Pijnenburg, G.~Cusin, and C.~Bonvin, \ifthenelse{\boolean{articletitles}}{\emph{{The impact of large-scale galaxy clustering on the variance of the Hellings-Downs correlation: numerical results}}, }{}\href{http://arxiv.org/abs/2411.08744}{{\normalfont\ttfamily arXiv:2411.08744}}\relax
\mciteBstWouldAddEndPuncttrue
\mciteSetBstMidEndSepPunct{\mcitedefaultmidpunct}
{\mcitedefaultendpunct}{\mcitedefaultseppunct}\relax
\EndOfBibitem
\bibitem{Allen:2024mtn}
B.~Allen, D.~Agarwal, J.~D. Romano, and S.~Valtolina, \ifthenelse{\boolean{articletitles}}{\emph{{Source anisotropies and pulsar timing arrays}}, }{}\href{http://arxiv.org/abs/2406.16031}{{\normalfont\ttfamily arXiv:2406.16031}}\relax
\mciteBstWouldAddEndPuncttrue
\mciteSetBstMidEndSepPunct{\mcitedefaultmidpunct}
{\mcitedefaultendpunct}{\mcitedefaultseppunct}\relax
\EndOfBibitem
\bibitem{Maggiore:1900zz}
M.~Maggiore, {\em {Gravitational Waves. Vol. 1: Theory and Experiments}}, Oxford Master Series in Physics, Oxford University Press, 2007\relax
\mciteBstWouldAddEndPuncttrue
\mciteSetBstMidEndSepPunct{\mcitedefaultmidpunct}
{\mcitedefaultendpunct}{\mcitedefaultseppunct}\relax
\EndOfBibitem
\bibitem{Armendariz-Picon:2005lfa}
C.~Armendariz-Picon, \ifthenelse{\boolean{articletitles}}{\emph{{Footprints of statistical anisotropies}}, }{}\href{https://doi.org/10.1088/1475-7516/2006/03/002}{JCAP \textbf{03} (2006) 002}, \href{http://arxiv.org/abs/astro-ph/0509893}{{\normalfont\ttfamily arXiv:astro-ph/0509893}}\relax
\mciteBstWouldAddEndPuncttrue
\mciteSetBstMidEndSepPunct{\mcitedefaultmidpunct}
{\mcitedefaultendpunct}{\mcitedefaultseppunct}\relax
\EndOfBibitem
\bibitem{Planck:2015igc}
Planck, P.~A.~R. Ade {\em et~al.}, \ifthenelse{\boolean{articletitles}}{\emph{{Planck 2015 results. XVI. Isotropy and statistics of the CMB}}, }{}\href{https://doi.org/10.1051/0004-6361/201526681}{Astron.\ Astrophys.\  \textbf{594} (2016) A16}, \href{http://arxiv.org/abs/1506.07135}{{\normalfont\ttfamily arXiv:1506.07135}}\relax
\mciteBstWouldAddEndPuncttrue
\mciteSetBstMidEndSepPunct{\mcitedefaultmidpunct}
{\mcitedefaultendpunct}{\mcitedefaultseppunct}\relax
\EndOfBibitem
\bibitem{Hajian:2003qq}
A.~Hajian and T.~Souradeep, \ifthenelse{\boolean{articletitles}}{\emph{{Measuring statistical isotropy of the CMB anisotropy}}, }{}\href{https://doi.org/10.1086/379757}{Astrophys.\ J.\ Lett.\  \textbf{597} (2003) L5}, \href{http://arxiv.org/abs/astro-ph/0308001}{{\normalfont\ttfamily arXiv:astro-ph/0308001}}\relax
\mciteBstWouldAddEndPuncttrue
\mciteSetBstMidEndSepPunct{\mcitedefaultmidpunct}
{\mcitedefaultendpunct}{\mcitedefaultseppunct}\relax
\EndOfBibitem
\bibitem{2012reco.book.....E}
G.~F.~R. {Ellis}, R.~{Maartens}, and M.~A.~H. {MacCallum}, {\em {Relativistic Cosmology}}, 2012\relax
\mciteBstWouldAddEndPuncttrue
\mciteSetBstMidEndSepPunct{\mcitedefaultmidpunct}
{\mcitedefaultendpunct}{\mcitedefaultseppunct}\relax
\EndOfBibitem
\bibitem{Lesgourgues:2011re}
J.~Lesgourgues, \ifthenelse{\boolean{articletitles}}{\emph{{The Cosmic Linear Anisotropy Solving System (CLASS) I: Overview}}, }{}\href{http://arxiv.org/abs/1104.2932}{{\normalfont\ttfamily arXiv:1104.2932}}\relax
\mciteBstWouldAddEndPuncttrue
\mciteSetBstMidEndSepPunct{\mcitedefaultmidpunct}
{\mcitedefaultendpunct}{\mcitedefaultseppunct}\relax
\EndOfBibitem
\bibitem{Jenkins:2018lvb}
A.~C. Jenkins and M.~Sakellariadou, \ifthenelse{\boolean{articletitles}}{\emph{{Anisotropies in the stochastic gravitational-wave background: Formalism and the cosmic string case}}, }{}\href{https://doi.org/10.1103/PhysRevD.98.063509}{Phys.\ Rev.\  \textbf{D98} (2018) 063509}, \href{http://arxiv.org/abs/1802.06046}{{\normalfont\ttfamily arXiv:1802.06046}}\relax
\mciteBstWouldAddEndPuncttrue
\mciteSetBstMidEndSepPunct{\mcitedefaultmidpunct}
{\mcitedefaultendpunct}{\mcitedefaultseppunct}\relax
\EndOfBibitem
\bibitem{Mingarelli:2014xfa}
C.~M.~F. Mingarelli and T.~Sidery, \ifthenelse{\boolean{articletitles}}{\emph{{Effect of small interpulsar distances in stochastic gravitational wave background searches with pulsar timing arrays}}, }{}\href{https://doi.org/10.1103/PhysRevD.90.062011}{Phys.\ Rev.\ D \textbf{90} (2014) 062011}, \href{http://arxiv.org/abs/1408.6840}{{\normalfont\ttfamily arXiv:1408.6840}}\relax
\mciteBstWouldAddEndPuncttrue
\mciteSetBstMidEndSepPunct{\mcitedefaultmidpunct}
{\mcitedefaultendpunct}{\mcitedefaultseppunct}\relax
\EndOfBibitem
\bibitem{limber1953analysis}
D.~N. Limber, \ifthenelse{\boolean{articletitles}}{\emph{The analysis of counts of the extragalactic nebulae in terms of a fluctuating density field.}, }{}Astrophysical Journal, vol.\ 117, p.\ 134 \textbf{117} (1953) 134\relax
\mciteBstWouldAddEndPuncttrue
\mciteSetBstMidEndSepPunct{\mcitedefaultmidpunct}
{\mcitedefaultendpunct}{\mcitedefaultseppunct}\relax
\EndOfBibitem
\bibitem{Bartelmann:1999yn}
M.~Bartelmann and P.~Schneider, \ifthenelse{\boolean{articletitles}}{\emph{{Weak gravitational lensing}}, }{}\href{https://doi.org/10.1016/S0370-1573(00)00082-X}{Phys.\ Rept.\  \textbf{340} (2001) 291}, \href{http://arxiv.org/abs/astro-ph/9912508}{{\normalfont\ttfamily arXiv:astro-ph/9912508}}\relax
\mciteBstWouldAddEndPuncttrue
\mciteSetBstMidEndSepPunct{\mcitedefaultmidpunct}
{\mcitedefaultendpunct}{\mcitedefaultseppunct}\relax
\EndOfBibitem
\bibitem{babak2024forecasting}
S.~Babak, M.~Falxa, G.~Franciolini, and M.~Pieroni, \ifthenelse{\boolean{articletitles}}{\emph{Forecasting the sensitivity of pulsar timing arrays to gravitational wave backgrounds}, }{} 2024\relax
\mciteBstWouldAddEndPuncttrue
\mciteSetBstMidEndSepPunct{\mcitedefaultmidpunct}
{\mcitedefaultendpunct}{\mcitedefaultseppunct}\relax
\EndOfBibitem
\bibitem{liang2023test}
Q.~Liang, M.-X. Lin, and M.~Trodden, \ifthenelse{\boolean{articletitles}}{\emph{A test of gravity with pulsar timing arrays}, }{}Journal of Cosmology and Astroparticle Physics \textbf{2023} (2023) 042\relax
\mciteBstWouldAddEndPuncttrue
\mciteSetBstMidEndSepPunct{\mcitedefaultmidpunct}
{\mcitedefaultendpunct}{\mcitedefaultseppunct}\relax
\EndOfBibitem
\bibitem{omiya2023hellings}
H.~Omiya, K.~Nomura, and J.~Soda, \ifthenelse{\boolean{articletitles}}{\emph{Hellings-downs curve deformed by ultralight vector dark matter}, }{}Physical Review D \textbf{108} (2023) 104006\relax
\mciteBstWouldAddEndPuncttrue
\mciteSetBstMidEndSepPunct{\mcitedefaultmidpunct}
{\mcitedefaultendpunct}{\mcitedefaultseppunct}\relax
\EndOfBibitem
\end{mcitethebibliography}

\appendix

\section{Positivity of the clustering contribution to the HD variance}
\label{sec:Positivity}
Intuitively, we expect the fact that we can observe only one universe, i.e.~one realization of the galaxy density field, to increase the deviation from the HD curve that one measures in a real observation. Indeed, the positivity of $\angbr{\rho^2_{ab}}^\clust$ can be shown  by writing the wave random field $h_A$ as a superposition of two independent  random fields: the background field $h_A^\st$ {corresponding to} the absence of anisotropies, and a perturbation $\delta h_A$ due to the presence of {anisotropies in the galaxy density}. More precisely, we introduce
\begin{equation}
    h_A(f, \bn) = h_A^\st(f, \bn) +\delta h_A(f, \bn)\,,
\end{equation}
and the ensemble averages over realizations of $h_A$ are treated as joint averages over realizations of both $h^\st_A$ and $\delta h_A$. For any realization $\rho_g$ of the density field, the means of background and perturbation wave fields are given by
\begin{equation}
    \angbr{ h_A^\st}_{h|\rho_g} = \angbr{ h_A^\st}_{(h^\st, \delta h)|\rho_g} = 0\,, \qquad \angbr{ \delta h_A}_{h|\rho_g} = \angbr{ \delta h_A}_{(h^\st, \delta h)|\rho_g} = 0\,, \label{eq:means}
\end{equation}
and the correlations by
\beq
\left\langle h_A^{\st \ast}(f,\mathbf{n})h^\st_B(f',\mathbf{n'})\right\rangle_{h|\rho_g} =\frac 12 \delta_{AB}\,\delta(f-f')\bar S_h(f)\frac{\delta^2(\mathbf{n},\mathbf{n'})}{4\pi}\,,
\eeq
and
\begin{align}
\left\langle \delta h_A^\ast(f,\mathbf{n})\delta h_B(f',\mathbf{n'})\right\rangle_{h|\rho_g} &=\frac 12 \delta_{AB}\,\delta(f-f') \delta S_h(f,\bn)\frac{\delta^2(\mathbf{n},\mathbf{n'})}{4\pi} \nonumber \\
&=  \frac 12 \delta_{AB}\,\delta(f-f')\bar S_h(f) \int\mathrm d z\,b_{\rm GW}(f,z)\delta_g(\mathbf{n},z) \frac{\delta^2(\mathbf{n},\mathbf{n'})}{4\pi}\,. \label{eq:dh_2point}
\end{align}
Moreover, the independence of $(h_A^\st, \delta h_A)$ makes any average over their product vanish,
\beq
\left\langle \delta h_A(f,\mathbf{n})h^\st_B(f',\mathbf{n'})\right\rangle_{h|\rho_g} = \left\langle \delta h_A(f,\mathbf{n})\right\rangle_{h|\rho_g} \left\langle h^\st_B(f',\mathbf{n'})\right\rangle_{h|\rho_g} =0\,.
\label{FieldsIndep}
\eeq
Each of the field $h_A^{\rm st.}$ and $\delta h_A$ is, for a given density realization $\rho_g$ of the universe, considered to be Gaussian, i.e.~fully specified by Eqs.~\eqref{eq:means}--\eqref{eq:dh_2point}. With these properties, the full field $h_A$ is as well Gaussian for each density field realization $\rho_g$, and obeys
\beq
\left\langle h_A(f,\mathbf{n})\right\rangle_{h|\rho_g} =0,  \qquad
\left\langle h_A^\ast(f,\mathbf{n})h_B(f',\mathbf{n'})\right\rangle_{h|\rho_g} =\frac 12 \delta_{AB}\,\delta(f-f') S_h(f,\bn)\frac{\delta^2(\mathbf{n},\mathbf{n'})}{4\pi}\,. 
\eeq
Thus, the expectation value  $\angbr{\rho_{ab}}=\big\langle\angbr{\rho_{ab}}\cond\big\rangle_{\pg}$ remains unchanged and equals the HD curve predicted in Ref.~\cite{Hellings:1983fr}.

Let us now consider $\rho_{ab}^2$, the quantity needed to calculate the variance of the HD correlation. Splitting the wave field as $h = h^\st+\delta h$ leads to $\rho_{ab}^2$ containing the following types of terms (with arguments suppressed for conciseness):
\begin{align}
\rho_{ab}^2 \sim \ &\mathcal{O}(h_A^{\st\ast} h^\st_{A'} h^\st_{A''} h_{A'''}^{\st\ast}) +  \mathcal{O}(\delta h_A^\ast \delta h_{A'} \delta h_{A''} \delta h_{A'''}^\ast) \nonumber \\ &+ \text{Terms$( h^\st \cdot \delta h\cdot \delta h\cdot \delta h)$} + \text{Terms$( \delta h \cdot  h^\st\cdot  h^\st\cdot h^\st)$} + \text{Terms$( \delta h \cdot \delta h\cdot  h^\st\cdot h^\st)$}\,.
\label{rho2decomposition}
\end{align}
When taking the $\left\langle \cdot \right\rangle_{h|\pg}$ ensemble average, the Terms$( h^\st \cdot \delta h\cdot \delta h\cdot \delta h)$ and Terms$( \delta h \cdot  h^\st\cdot  h^\st\cdot h^\st)$ vanish in virtue of Eq.~\eqref{FieldsIndep}, as they are linear in either $h_A$ or $\delta h_A$ which are zero-mean random fields. The Terms$( \delta h \cdot \delta h\cdot  h^\st\cdot h^\st)$ do not vanish immediately when taking the $\left\langle \cdot \right\rangle_{h|\pg}$ ensemble average, but given Eq.~\eqref{eq:dh_2point} it leads to an expression linear in $\delta_g$, which vanishes once averaging over density field realizations to obtain the full ensemble average.

Finally, one recovers the results of section~\ref{Sec:HDvariance} for $\angbr{\rho_{ab}^2}^\st$ and $\angbr{\rho_{ab}^2}^\clust$, respectively, when taking the full ensemble average over the terms in the first line of Eq.~\eqref{rho2decomposition}. We thus still have $\angbr{\rho_{ab}^2} = \angbr{\rho_{ab}^2}^\st+\angbr{\rho_{ab}^2}^\clust$.
However, splitting $h_A$ into the fields $h_A^\st$ and $\delta h_A$ allows to define the random variable
\begin{equation}
 \Tilde{\rho}_{ab}   =\sum_{A,A'}\int\mathrm df\int\mathrm df'\int\mathrm d\bn\int\mathrm d\bn' R^A_a(f,\bn)^\ast R_b ^{A'}(f',\bn')\delta h_A^\ast(f,\bn)\delta h_{A'}(f',\bn')\mbox{sinc}\rbr{\pi(f-f')T}\,,
\end{equation}
i.e.~the same as the definition of $\rho_{ab}$ given in Eq.~\eqref{Eq:rho_ab}, but with the perturbation $\delta h_A$ instead of the full $h_A$ (beware that $\rho_{ab} \neq \rho^\st_{ab} + \Tilde{\rho}_{ab}$, since $\rho_{ab}$ is quadratic in the strain). This variable satisfies $\angbr{\Tilde{\rho}_{ab}}= 0$, again due to the fact that the ensemble average $\left\langle \cdot \right\rangle_{h|\pg}$ leads to an expression linear in $\delta_g$.
Therefore, $\text{Var}[\Tilde{\rho}_{ab}] = \angbr{\Tilde{\rho}^2_{ab}} \geq 0$.
The evaluation of $\angbr{\Tilde{\rho}^2_{ab}}$ follows the exact same steps as that of $\angbr{\rho_{ab}^2}$ in the absence of clustering, substituting $h_A^\st$ by $\delta h_A$. In particular, we can still use Isserlis' theorem since $\delta h_A$ is Gaussian for a given density realization of the universe. We immediately obtain
\beq
\text{Var}[\Tilde{\rho}_{ab}] =\angbr{\Tilde{\rho}^2_{ab}}= \angbr{\rho_{ab}^2}^\clust \geq0\,.
\eeq

\end{document}